# Deep-level engineering in aluminum nitride (AlN) enables reliable photodetection in extreme conditions

Jiahao Dong[1], Zhenjing Liu[1], Sara Tulchinsky[1], Rafael Jaramillo[1]*

1. Department of Materials Science and Engineering, Massachusetts Institute of Technology; Cambridge, Massachusetts 02139, USA.

*Corresponding author: rjaramil@mit.edu

**Abstract:** Semiconductor photodetectors are conventionally optimized for sensing weak optical signals, and they typically saturate at low-to-moderate light intensity. Here, we demonstrate AlN photodetectors that exhibit non-saturating linear response to ultra-bright, sub-bandgap light exceeding 40 W/cm$^2$ (over 400 Suns). The photodetector further shows undistorted linear response at elevated temperature, up to at least 300 ºC. This exceptional performance originates from photoresponse mediated by point defects with energy levels deep in the bandgap (deep levels) at the metal-AlN Schottky junction. Through dopant design and contact engineering, we demonstrate that a narrow space-charge region is essential for enabling ultra-bright light detection and accurate measurement. These results establish a strategy for engineering ultra-wide bandgap (UWBG) semiconductor devices for reliable operation in extreme conditions to meet emerging needs in industrial process control, thermal and nuclear power generation, and aeronautics and spaceflight.

## Main Text:

Accurate detection and imaging of intense light are desired for emerging technologies including laser-based manufacturing, plasma and fusion systems, advanced combustion engines, and aerospace and defense platforms.[1–3] Nonetheless, semiconductor photodetectors have been developed primarily for detection of weak optical signals, with figures of merit such as detectivity and noise-equivalent power (NEP) optimized for sensitivity at low optical power.[4] Under intense (*i.e.*, high-irradiance) illumination, conventional semiconductor photodetectors typically become nonlinear and may ultimately fail because of trap filling, space-charge effects, carrier transport limitations, and overheating.[5–9] This nonlinearity impedes accurate measurements of irradiance and results in loss of quantitative information. Despite advances in photodetector technologies, semiconductor devices capable of directly and quantitatively sensing bright light remain elusive.

Here, we report a defect-engineered aluminum nitride (AlN) photodetector that features linear, non-saturating response to ultra-bright sub-bandgap illumination exceeding 40 W/cm$^2$, and maintains undistorted operation up to at least 300 °C. These performance levels substantially exceed those of existing semiconductor photodetectors. Sub-bandgap sensitivity results from introducing a point defect (here, Ge) with an energy level deep in the AlN bandgap. Although defect-mediated photoconductivity has long been observed in semiconductors, it is generally considered unsuitable for photodetection because the finite population of defect states can be fully depleted under intense illumination, causing saturated photoresponse.[10–15] Here, we overturn this conventional expectation and demonstrate that non-saturating sub-bandgap response can be achieved through tunneling-based defect recombination at a suitably-designed metal-AlN heterojunction. This mechanism enables robust optical sensing across operating regimes that are inaccessible to existing semiconductor photodetectors.

These results reveal a route to new optoelectronic functionality in ultra-wide-bandgap (UWBG) semiconductors through the deliberate control of deep-level defects. By overcoming the long-assumed performance limit of defect-mediated photoconductivity, this work establishes a framework for treating defect states as active device elements rather than merely as material imperfections. Because deep-level defects are ubiquitous across WBG and UWBG semiconductors, the underlying principles demonstrated here may be broadly applicable to a wide range of materials and device architectures.

### Device concept and performance

We fabricate photodetectors in a lateral configuration, with metal contacts (1 nm Ti / 80 nm Al) deposited on an epitaxial AlN thin film grown by metal-organic chemical vapor deposition (MOCVD, Adroit Materials) with a nominal Ge concentration of $2\times10^{19}$ cm$^{-3}$. The contacts are square pads measuring 300 × 300 μm, separated by 50 μm. Ge is expected to be a deep level in AlN because of its high thermal ionization energy.[16,17] Consistently, the AlN:Ge films are found semi-insulating. To expose the device to high-irradiance light, we use a two-lens optical system to focus light from a blue LED (Thorlabs M450LP2 with a center wavelength 450 nm, **Fig. S1**).

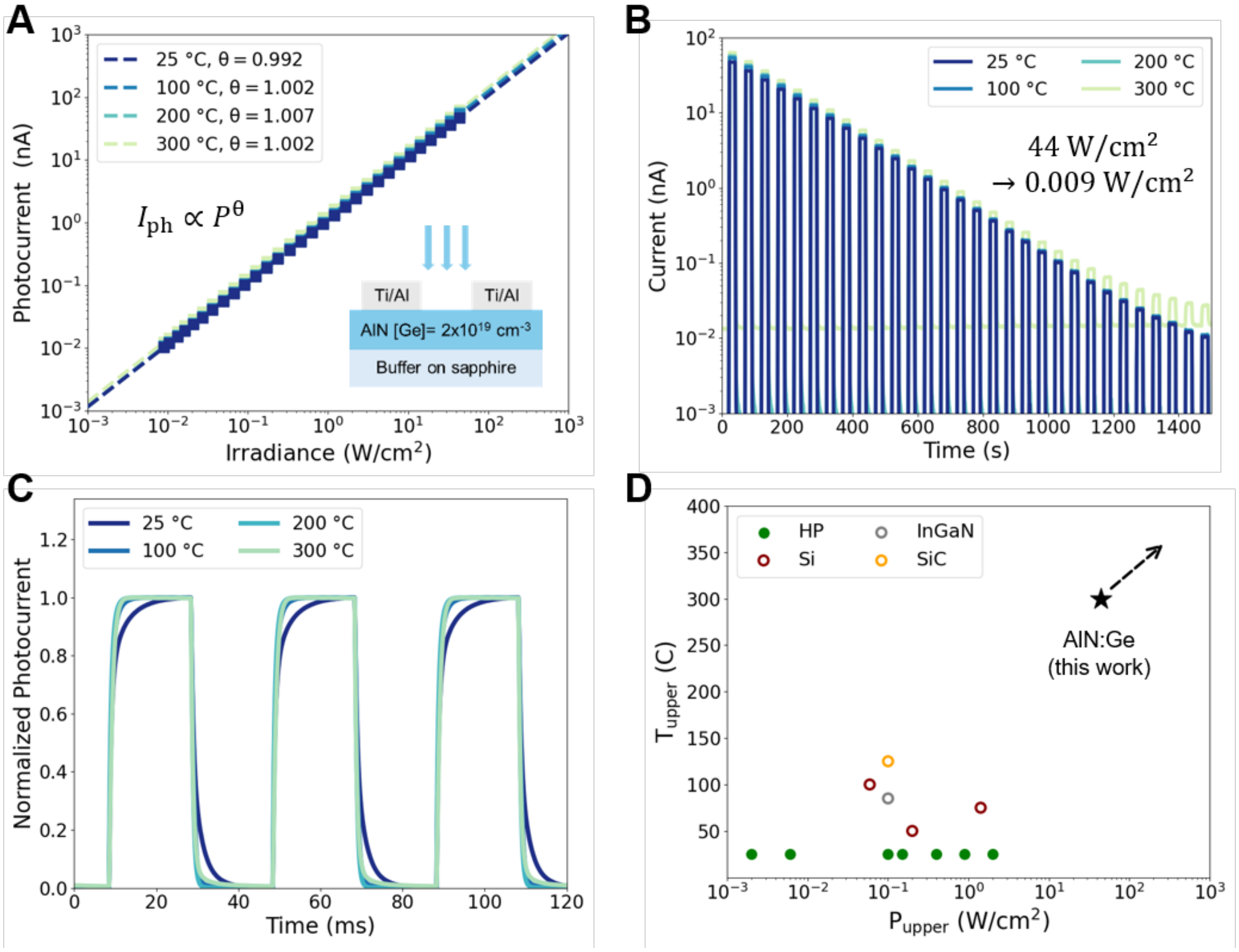

**Figure 1: Linear and stable photoresponse at high irradiance and elevated temperatures.** (**A**) Photocurrent ($I_{\text{ph}}$) *vs*. irradiance ($P$) measured between 25 – 300 °C. The dashed lines are power-law fits, and dark current levels have been subtracted. All data in panels (A-C) are recorded at a DC voltage of 5 V. Inset: Device structure (not to scale) and incident illumination. (**B**) Current timeseries data as illumination is cycled on and off in decreasing steps from 44 – 0.009 W/cm$^2$. (**C**) Normalized transient photoresponse under irradiance of 44 W/cm$^2$. (**D**) Comparison of the maximum demonstrated irradiance and operating temperature of our device with the upper limits of previously reported photodetectors at comparable wavelengths, including commercial products (open symbols) and literature reports (filled symbols), all operating in the linear response regime (see Data S1 for underlying data). HP = halide perovskites.

In **Fig. 1A** we present the device photocurrent ($I_{\text{ph}}$) *vs.* irradiance ($P$) measured between room temperature and 300 °C. We model the photoresponse as a power law $I_{\text{ph}} \propto P^{\theta}$ and fit the scaling exponent θ at each temperature. According to the consensus statement by Pecunia *et al.*, a photodetector is classified as linear when the exponent θ lies within the range 0.99 - 1.01.[4] The data and fits presented in **Fig. 1A** confirm that the AlN:Ge photodetector maintains a non-saturating, linear response under blue illumination up to 44 W/cm$^2$ and temperatures up to 300 °C. These are the limits of our test equipment; the upper limits of the device may be higher. In **Fig. 1B** we present a photocurrent time series for cyclic illumination with decreasing irradiance. The dark current is low at all measurement temperatures, due to the UWBG of AlN and the high ionization energy of Ge deep levels. As a result, the ON/OFF ratio is high at all test conditions.

In **Fig. 1C** we present the transient photoresponse of the AlN:Ge device. At 25 °C, both the 90% rise time and 90% fall time are approximately 4 ms, decreasing to 1.5 ms at 300 °C.

Importantly, the AlN:Ge device exhibits excellent operational stability, exhibiting no measurable degradation in photoresponse after 100,000 illumination cycles (20 ms on/20 ms off) at 44 W/cm$^2$. We compare the irradiance-dependent photoresponse and transient behavior of the AlN:Ge device with those of commercial Si and InGaN photodiodes (**Fig. S2**). Under bright illumination, the commercial devices are nonlinear and show performance degradation, whereas the AlN:Ge device maintains a stable photoresponse during prolonged exposure. The spectral range of the AlN:Ge photodetector extends over at least 360 - 490 nm (**Fig. S3**), spanning the ultraviolet (UV) to blue region.

In **Fig. 1D** we benchmark the maximum demonstrated irradiance and operating temperature of our device with the upper limits of previously reported photodetectors at comparable wavelengths, including commercial products and literature reports, all operating in the linear response regime.[18–23] The AlN:Ge photodetector presented here exceeds prior reports by significant margins in irradiance and operating temperature. Although reports of photodetectors for high temperature operation based on WBG and UWBG semiconductors have increased in frequency in recent years, linear response was not achieved at elevated temperatures.[24–29]

**Device operating principle**

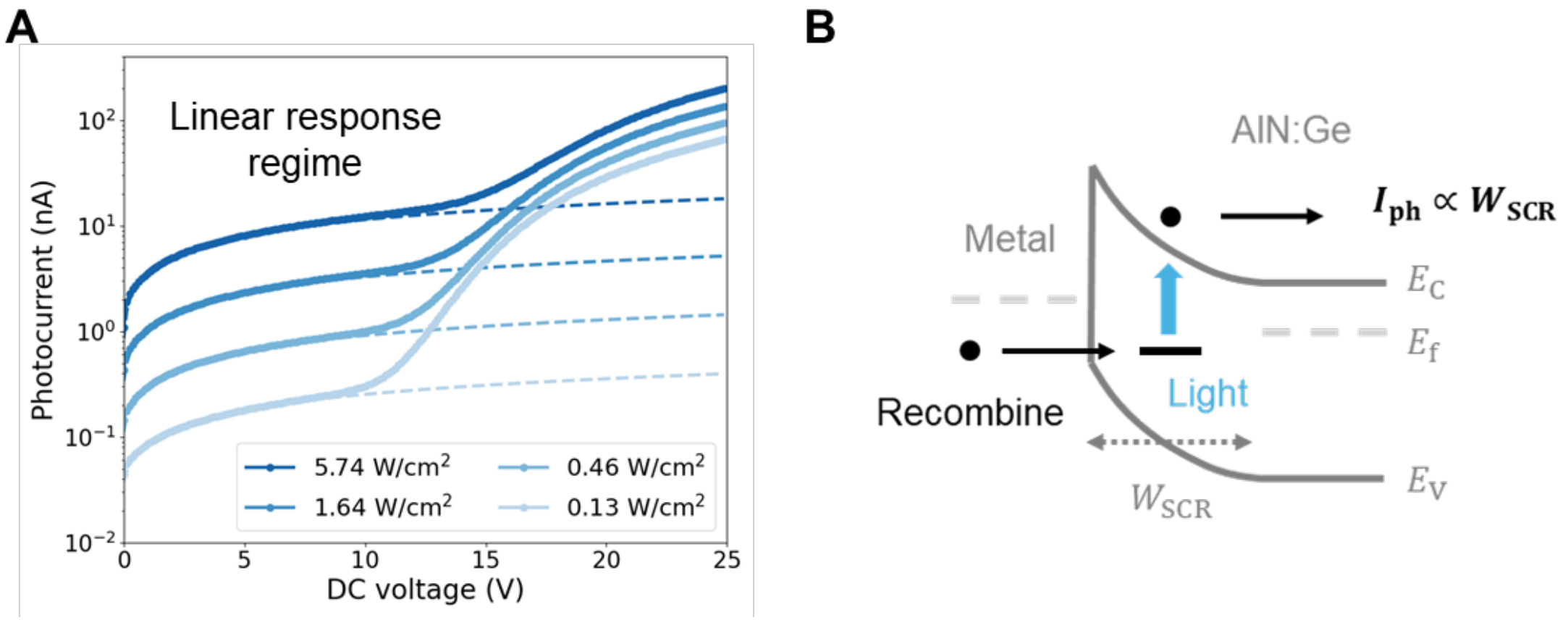

**Figure 2: Relationship between linear photoresponse and $W_{SCR}$.** (**A**) Photocurrent ($I_{ph}$) *vs.* DC voltage ($V_{DC}$) at discrete irradiance levels; dark current levels have been subtracted. Linear response is observed below approximately 10 V. Dashed lines are best fits of the data to the form $I_{ph} \propto \sqrt{V_{DC} + \phi}$, where $\phi$ a fitting parameter (see table S1 for best-fit values). (**B**) Schematic band diagram illustrating defect ionization by sub-bandgap photoexcitation, and recombination by tunneling injection from the metal contact. Photoexcited carriers drift in the built-in electric field and contribute to $I_{ph}$.

To understand the mechanism behind the non-saturating response of our AlN:Ge device, we measured photocurrent $I_{ph}$ *vs.* DC voltage $V_{DC}$ for varying irradiance (**Fig. 2A**). Linear response is achieved below approximately 10 V; at higher voltage, $I_{ph}$ increases rapidly and exhibits sub-linear dependence on the irradiance. In the linear response regime, the dependence of $I_{ph}$ on $V_{DC}$ can be well fit to the form $I_{ph} \propto \sqrt{V_{DC} + \phi}$, where $\phi$ is a fitting parameter, demonstrated by the dashed lines in **Fig. 2A**. For an applied DC voltage, one of the metal-AlN junctions is under reverse bias $V_R \approx V_{DC}$, while the other junction is under near-zero bias.[14] The width of the space charge region (SCR) $W_{SCR}$ at a Schottky junction follows a square root dependence on the reverse bias $V_R$, *i.e.*, $W_{SCR} \propto \sqrt{V_R + \phi_{bi}}$, with $\phi_{bi}$ the built-in potential.[30] This suggests that the photocurrent is likely proportional to the space charge region width, $I_{ph} \propto W_{SCR}$.

We further propose that the $I_{\mathrm{ph}}$ in the linear response regime originates from defect-to-conduction band excitation within the SCR, as illustrated in **Fig. 2B**. When exposed to sub-bandgap light, photocarriers generated via defect excitation are swept out of the SCR by the internal electric field, contributing to the photocurrent $I_{\mathrm{ph}}$. Besides, a non-saturating response at high irradiance requires that the recombination (carrier recapture) at the photoionized defect states is sufficiently fast, so that the defect states do not become exhausted under illumination.[12] Nonetheless, within the SCR, free carriers are depleted, and there are vanishingly few conduction electrons available for defect recombination.[31] This motivates our hypothesis that carrier injection by tunneling from the metal contact provides sufficiently fast recombination and enables non-saturating photoresponse. Defect-to-band tunneling rates become appreciable for internal electric field strength exceeding $10^5$ V/cm,[32,33] which is achieved at the reverse biased junction for $V_{\mathrm{DC}}$ below 10 V (see below).

The mechanism discussed above implies that achieving linear photoresponse requires the presence of a narrow SCR at the reversed biased metal-AlN junction. Device performance should suffer if the SCR is eliminated by forming an ohmic contact, or if the SCR becomes too wide to enable tunneling injection. We test these predictions by designing devices with suppressed and widened SCRs. The SCR can be suppressed by thermal annealing of the metal contact at 950 °C, which promotes interdiffusion at the metal-nitride interface and converts the contact from Schottky to near-ohmic behavior.[34,35] The SCR can be widened by lowering the Ge concentration, as $W_{\mathrm{SCR}}$ scales inversely with the square root of the dopant concentration.[30] Accordingly, we fabricate and test an AlN photodetector with a nominal Ge incorporation of $2\times10^{18}$ cm$^{-3}$.

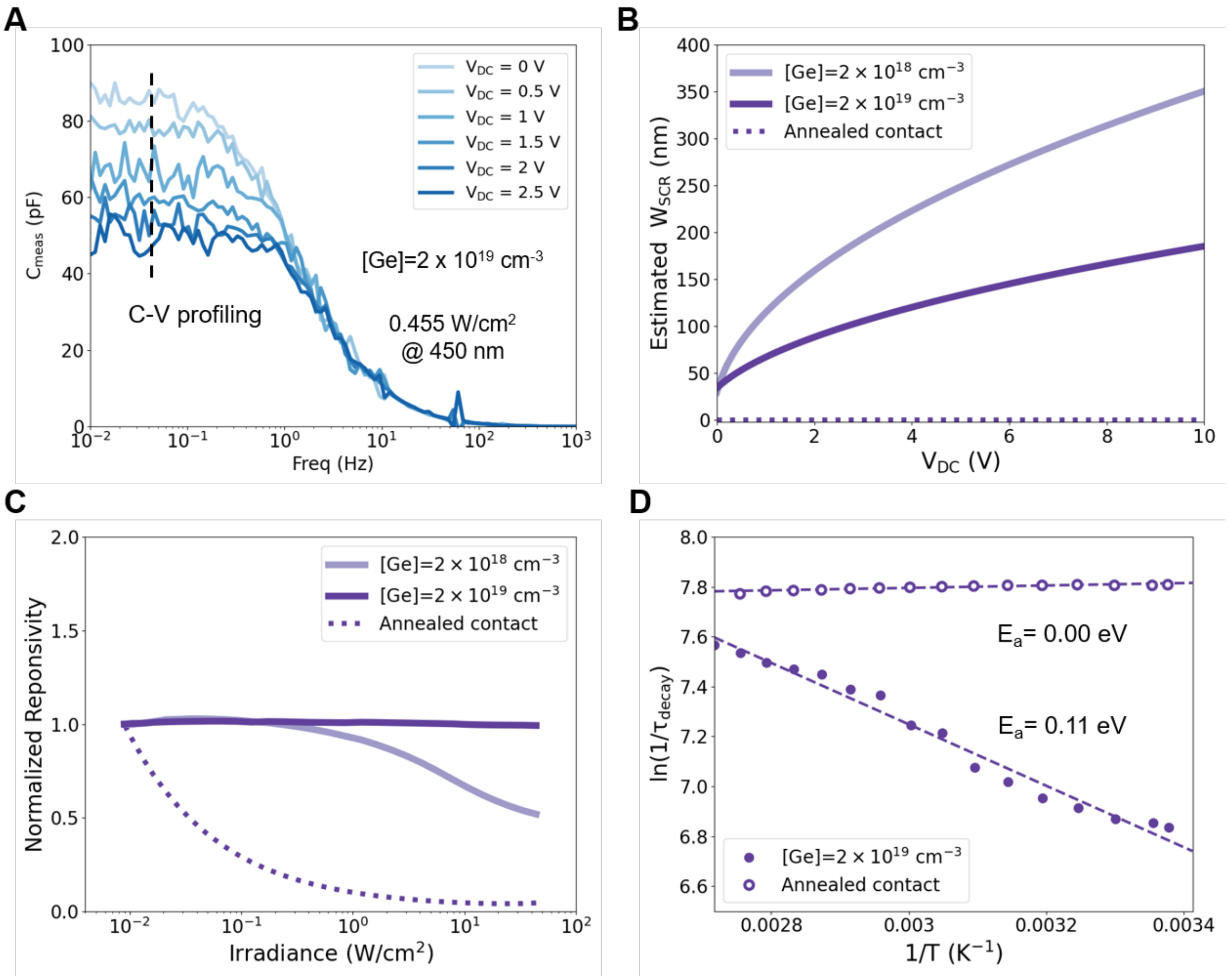

**Figure 3: Illustrating the importance of a narrow SCR for device performance.** (**A**) *C-V* profiling data for device with [Ge] = $2\times10^{19}$ cm$^{-3}$, measured with irradiance of 0.455 W/cm$^2$, for $V_{DC}$ between 0 and 2.5 V. The dashed line indicates the low-frequency data used for Mott-Schottky analysis. (**B**) $W_{SCR}$ *vs.* $V_{DC}$ for three devices with different doping and/or processing. The solid lines represent devices with non-annealed contacts and different [Ge]. The dashed line represents a device with annealed contacts and [Ge] = $2\times10^{19}$ cm$^{-3}$. (**C**) Optical responsivity for irradiance from 0.009 - 44 W/cm$^2$ for the same three devices presented in (A). The data are normalized to the responsivity measured at the lowest irradiance (0.009 W/cm$^2$), to illustrate linearity and deviation from linearity. (**D**) Arrhenius analysis of photocurrent decay with the activation energy $E_a$ extracted for the temperature range from 23 to 90 °C.

To determine $W_{SCR}$, we use capacitance-voltage (*C-V*) profiling and Mott-Schottky analysis to extract the effective defect concentration and estimate $W_{SCR}$. Due to the very low free carrier concentrations in the AlN:Ge, and the inability of deep defects to respond to high frequency modulation (*e.g.*, MHz), low-frequency *C-V* measurements are needed to properly extract $W_{SCR}$.[36–38] *C-V* profiling data for the device with [Ge] = $2\times10^{19}$ cm$^{-3}$ are shown in **Fig. 3A**; further details are presented in Supplementary Information and **Fig. S4**. The extracted $W_{SCR}$ are presented in **Fig. 3B**. The junction with a nominal defect concentration [Ge] = $2\times10^{19}$ cm$^{-3}$ shows a relatively narrow SCR: at $V_{DC} = 5$ V, $W_{SCR} \approx 132$ nm and the peak electric field approaches $7.5\times10^5$ V/cm. After contact annealing, the junction shows near-ohmic behavior (**Fig. S5**) and the SCR has collapsed to $W_{SCR} \approx 0$ .[34] Cross-sectional imaging of the annealed contact with scanning transmission electron microscopy (STEM) confirms that annealing-induced interdiffusion is confined to within 2 nm of the metal-AlN interface, without noticeable modification of the AlN bulk (**Fig. S6**). Meanwhile, lowering [Ge] from $2 \times 10^{19}$ cm$^{-3}$ to $2 \times 10^{18}$ cm$^{-3}$ results in a wider SCR, as expected.

In **Fig. 3C** we present the photoresponse of these three devices: one with a narrower SCR ([Ge] = $2 \times 10^{19}$ cm$^{-3}$), one with a wider SCR ([Ge] = $2 \times 10^{18}$ cm$^{-3}$), and one with a negligible SCR (contact annealed). We present the normalized optical responsivity *vs.* irradiance to assess linearity. The [Ge] = $2 \times 10^{19}$ cm$^{-3}$ device exhibits constant responsivity and thereby linear photoresponse. In contrast, the [Ge] = $2 \times 10^{18}$ cm$^{-3}$ device exhibits response saturation, consistent with exhaustion of defect levels. The annealed contact device shows rapid performance loss with increasing irradiance. These results illustrate that linear, non-saturating photoresponse depends on the existence of a sufficiently narrow SCR.

The SCR determines device performance through the recombination mechanism, as indicated in **Fig. 2B**. The tunneling-based recombination pathway may be thermally-activated, as in thermionic field emission.[30] In contrast, the device with the annealed contact has no Schottky junction, and recombination relies primarily on the cross section for electron recapture from the conduction band, which may be barrier-free (or nearly so) and temperature-independent.[15] To examine the recombination mechanism, we measured photocurrent decay (*i.e.*, persistence) as a function of temperature and extracted the decay rates $1/\tau_{\mathrm{decay}}$ (**Fig. S7**), which are presented in the Arrhenius analysis in **Fig. 3D**. In the temperature range of 25 to 90 °C, the [Ge] = $2 \times 10^{19}$ cm$^{-3}$ device with non-annealed contacts shows an activation energy of 0.11 eV, whereas in the annealed contact device the decay rate is nearly temperature independent. These results confirm that the SCR determines recombination mechanism and thereby device performance. The measured thermal barrier of 0.11 eV in the photocurrent decay may be further lowered by optimizing the metal contact layer.

**Conclusion**

We demonstrate that deep-level-mediated photoconductivity can be exploited in a UWBG semiconductor to enable non-saturating response to ultra-bright sub-bandgap light. This runs contrary to decades of accepted wisdom about intrinsically poor performance of devices harnessing defect-mediated photoconductivity. The device presented based on Ge-doped AlN with [Ge] = $2 \times 10^{19}$ cm$^{-3}$ exhibits linear and robust response for irradiance as high as 44 W/cm$^2$ at a center wavelength of 450 nm, and for temperatures as high as 300 °C. Because of lack of free carriers in the AlN, the device exhibits negligible high-frequency capacitance. Thus, the Ge-doped AlN behaves as a low-capacitance current source upon sub-bandgap photoexcitation, which is promising for high-speed operation. Also, devices that rely on tunneling injection are well suited for high-speed switching owing to the intrinsically fast nature of quantum-mechanical tunneling.[39]

The operating principle demonstrated here expands the frontier of UWBG semiconductor optoelectronics. The spectral response to sub-bandgap light can be adjusted through dopant selection and/or alloy design; for instance, literature suggests that Si doping and using high-Al-content AlGaN alloys may enable near-infrared sensitivity.[16,40] Semiconductor photodetectors capable of linear measurement of high-irradiance light at high temperatures may find many applications in the industrial and defense sectors, including but not limited to laser- and plasma-based materials manufacturing processes, next-generation fusion and fission plants, military sensing and imaging applications, etc.